Magnetization Losses in Multiply Connected $YBa_2Cu_3O_{6+x}$ Coated Conductors.

G. A. Levin and P. N. Barnes

Propulsion Directorate, Air Force Research Laboratory, 1950 Fifth St. Bldg. 450, Wright-Patterson Air Force Base OH 45433

Naoyuki Amemiya, Satoshi Kasai, Keiji Yoda, and Zhenan Jiang

Faculty of Engineering, Yokohama National University,
79-5 Tokiwadai, Hodogaya, Yokohama 240-8501, Japan

A. Polyanskii
1500 Engineering Dr., ERB Applied Superconductivity Center, University of Wisconsin, Madison WI 53706

## Abstract.

We report the results of a magnetization losses study in experimental multifilament, multiply connected coated superconductors exposed to time-varying magnetic field. In these samples, the superconducting layer is divided into parallel stripes segregated by non-superconducting grooves. In order to facilitate the current sharing between the stripes and thus increase the reliability of the striated conductors, a sparse network of superconducting bridges is superimposed on the striated film. We find that the presence of the bridges does not substantially increase the magnetization losses, both hysteresis and coupling, as long as the number of bridges per length of the sample is not large. These results indicate that it is possible to find a reasonable compromise between the competing requirements of connectivity and loss reduction in an ac-tolerant version of the high temperature coated conductors specifically designed for ac power applications.

## 1. Introduction.

Replacement of copper wires with superconductors can result in substantial reductions in the size and weight of high power transformers, motors and generators[1-4]. The maximum benefit of such reduction can potentially be achieved in a fully superconducting (all-cryogenic) machine where both stator and rotor windings are superconducting. The main obstacle to the implementation of such designs is the very large ac loss that would occur in the superconducting armature winding because it is subjected to an alternating magnetic field [5-11].

This problem is especially acute in the second generation $YBa_2Cu_3O_{6+x}$ (YBCO) coated conductors that are produced in the form of wide, thin tapes [12-16]. A way to reduce the hysteresis loss in such tapes by dividing them into a large number of parallel stripes segregated by non-superconducting resistive barriers was proposed in Refs. [17,18]. An experiment on small samples of YBCO films deposited on LaAlO3 substrate has confirmed the validity of that suggestion [19].

Measurements of the magnetization losses in actual multifilamentary YBCO coated superconductors have also been reported recently[20-22]. In Refs.[20,22] two 1 x 10 cm samples of YBCO coated conductor, one control and the other striated, were subjected to a magnetic field normal to the wide face of the tape and varying at different linear frequencies $f$ in the range 11-170 Hz with a magnetic induction amplitude B up to 70 mT. The relatively long samples and measurements at several frequencies enabled the study of both hysteresis and coupling losses. The combined loss in the multifilament conductor was reduced by about 90% in comparison with the uniform conductor at full field penetration at sweep rate B$f$ as high as 3T/s [22].

This straightforward solution of the ac loss problem is not without serious drawbacks. One of them is lack of connectivity, i.e. the ability of superconducting stripes to share supercurrent should one of them become blocked, temporarily or permanently, by a localized defect or a "hot spot." The origin of defects may be the misalignment of the grains in thin superconducting films, as well as the inevitable defects appearing in the conductor during its manufacturing and exploitation - mechanical stresses, spontaneously overheated areas, etc. In a striated tape without current sharing between the stripes these defects lead to cumulative degradation of the current-carrying capacity of the long conductors. This problem will become more acute with increasing number density of filaments. In the non-striated wide tape such defects do not have the cumulative effect on the ability of the long conductor to carry transport current because the supercurrent can circumvent the damaged areas and recombine behind them.

The normal operating regime of a conductor in a generator, motor, or transformer is sub-critical (the transport current is well below critical). In the sub-critical regime the resistive connections between the stripes cannot facilitate the sharing of the transport current, because there is no potential difference between the stripes (the superconducting areas are equipotential). Therefore, the best way to achieve a degree of connectivity in the striated superconductor operating in the sub-critical regime is to provide a system of discrete or continuous superconducting bridges between the stripes [23].

However, such interconnections may increase the magnetization losses that depend on the pattern of penetration of the magnetic field. The currents induced across the superconducting bridges and their metal over-layer become an additional source of dissipation which is absent in the fully striated tapes. Therefore, a layout of the coated superconducting tapes for ac applications has to be devised as a compromise between competing requirements – connectivity on one hand and the reduction of the net loss on the other.

This paper is structured as follows: Section 2 describes the motivation behind the concept of multiply connected superconducting films for ac applications. In subsection (**A**) we consider two primary sources of losses in multifilament conductors – hysteresis and coupling – and show that the total loss at the operating sweep rate of the applied field can be reduced to the desired level by dividing the superconducting film into many parallel stripes. This, however, is predicated on the design of the conductor with correct correlation of such properties as the number density of stripes, the effective resistance of the barriers separating the superconducting stripes, and the length of the twist pitch. In subsection (**B**) the vulnerability of multifilament conductors to small defects with the size comparable to the width of an individual stripe is discussed. It is shown that without current sharing between the stripes the usable length of the conductor as a whole is a fraction of a defect-free length of an individual stripe. With decreasing width of the stripes the probability of blockage increases exponentially, which makes it highly unlikely that a multifilament conductor without current sharing can be manufactured in substantial length.

Sections 3 and 4 are devoted to the experimental results in which the ac losses were measured in several differently patterned striated coated superconductors: fully striated (simply connected) and two multiply connected versions. The fully striated samples are comprised of parallel superconducting stripes segregated by non-superconducting resistive barriers. They have the lowest losses, but the lack of connectivity makes such a pattern vulnerable to defects, especially with fine filaments.

Multiply connected patterns are devised as a compromise intended to ensure some degree of connectivity as well as low ac losses. Their important feature is that any two points of the superconducting film are connected by superconducting paths, but the supercurrent induced by alternating magnetic field is still predominantly channeled into narrow stripes. Magneto-optical imaging was used to obtain the overall picture of magnetic field penetration in both, fully striated and multiply connected samples.

## 2. Motivation for Multiply Connected Coated Superconductors.

### A) Losses in multifilament coated conductors.

The rate of Joule heat generation by an electric field $Q = \int \vec{j} \cdot \vec{E} dV$, where integration is extended over the volume of a conductor. Coated conductors are composite materials. In the superconducting film, the current density $j$ is essentially constant, equal to the critical value (Bean model), and its direction is determined by the direction of the electric field: $\vec{j}_s = j_c \vec{E}/|E|$. As the result, the amount of heat released in the superconducting layer is given by [5]

$$Q_s = j_c \int |\vec{E}| dV_s . \tag{1}$$

This component of the losses is usually called the hysteresis loss and the integration is carried out over the volume of the superconductor. In the normal metal the losses are given by

$$Q_n = \int \rho^{-1} |\vec{E}|^2 dV_n . \tag{2}$$

Here $|\vec{E}|$ and $|\vec{E}|^2$ are local quantities *averaged over the time cycle* and the resistivity $\rho$ may be nonuniform. Eq. (2) includes the integration over the volume of the substrate, as well as the stabilizer. Eqs. (1) and (2) have to be applied to a particular topology of the superconducting layer and coupled with the Maxwell equations. The reduction of losses in striated superconducting tapes is the result of reduction of the induced electric field which is determined by the pattern of penetration of the magnetic field.

The total power loss in a rectangular uniform superconducting film of width *W*, length *L* and negligible thickness, exposed to a harmonically time-varying magnetic field is given by [24]

$$Q_s \approx J_c W^2 L (B - B_c) f \ ; \ \ B >> B_c \tag{3}$$

Here $J_c$ is the density of critical current per unit width of the tape, $B_c = \mu_0 J_c \ln 4/\pi$, and $\mu_0 = 4\pi \times 10^{-7}$ H/m is the magnetic permeability of vacuum. The condition $B >> B_c$ defines the full penetration regime. For YBCO coated conductors the typical value of $J_c \approx 100$ A/(cm width) which translates into $B_c \approx 10$ mT. Hereafter we estimate losses only in the practically important regime of full penetration. We also will consider the power loss expressions given below to be normalized per unit length of the conductor.

In order to reduce the hysteresis loss the uniform superconducting film can be divided into parallel stripes[17]. In the multifilament superconductor comprised of N stripes the hysteresis loss is reduced in proportion to the width of an individual stripe $W_n$ [25]:

$$Q_s \approx I_c W_n (B - B'_c) f \ ; \quad B >> B'_c , \tag{4}$$

where $I_c = J_c N W_n$ is the total critical current and

$$B'_c = \frac{\mu_0 J_c \ln 4}{\pi} M(a) ;$$

$$M(a) = -\frac{1}{a^2 \ln 2} \int_0^\infty x dx \ln\left(1 - \frac{\sin^2(a)}{\cosh^2(x)}\right) ; \tag{5}$$

$$a = \pi W_n / 2\Delta W .$$

Here $\Delta W = W/N$ is the distance between the centers of the neighboring stripes. For example, if the number density of stripes per unit width is 20 stripes /cm, as is the case with most of the samples discussed below, $\Delta W = 0.5$ mm, while the width of the superconducting stripes $W_n \approx 0.47$ mm. In this case $W_n/\Delta W \approx 0.9$, which corresponds to $M(a) \approx 0.7$. In the limit of very wide grooves between the stripes, when $W_n/\Delta W << 1$, $M(a)$ tends to unity.

The purpose of striation, as it was originally envisioned [17], is to reduce the hysteresis loss in the full penetration regime by a factor $1/N$. However, in the multifilament conductor another channel of dissipation (coupling losses) appears as a result of the induced currents passing between the superconducting stripes through the normal barriers. In a rectangular flat multifilament sample of length L the electric field in the direction perpendicular to the superconducting stripes [5] is

$$E_\perp(x,t) \approx B \omega x \mathrm{Sin}(\omega t) , \tag{6}$$

where $-L/2 < x < L/2$ is the coordinate along the length of the sample. Assuming that the magnitude of the electric field is approximately uniform over the cross-section area of the substrate, we obtain from Eq.(2):

$$Q_n = \int \rho^{-1} |\vec{E}_\perp|^2 dV_n \approx \frac{\pi^2}{6} \frac{(BfL)^2}{\rho} d_n W . \tag{7}$$

Here $d_n$ is the thickness of the metal substrate. The coupling power loss per unit length increases quadratically with the length of the sample. Therefore, in order to limit this component of loss, a long multifilament conductor has to be twisted[17]. Equation (7) also describes the coupling loss in a twisted conductor, where *L* is half of the twist pitch. The numerical coefficient may be different depending on the type of the twist. Thus, at full field penetration the total magnetization power loss is determined by a simple quadratic function of the *sweep rate* $Bf$ (see Eqs. (4) and (7)):

$$Q = Q_s + Q_n \approx q_s Bf + q_n (Bf)^2; \quad q_s = W_n I_c; \quad q_n = \frac{\pi^2}{6} \frac{L^2}{\rho} d_n W. \quad (8)$$

The linear term describes the loss in the superconducting material and the quadratic term – the loss in the normal metal. The Eq. (8) can be rewritten in a more physically transparent form:

$$Q = \lambda_1 I_c Bf \left(1 + \frac{Bf}{\mathcal{R}}\right); \quad \lambda_1 \approx W_n; \quad \mathcal{R} = \frac{q_s}{q_n} \approx \frac{6}{\pi^2} \frac{W_n \rho I_c}{L^2 d_n W}. \quad (9)$$

Here, $\mathcal{R}$ is the *break-even sweep rate* at which the coupling loss is equal to the hysteresis loss. It determines approximately the optimal range of the operating sweep rate that the conductor should be exposed to in a particular application [22]. Thus, in order to reduce the total loss to acceptable level, one has to reduce the width of an individual stripe, while maintaining the value of the break-even rate $\mathcal{R}$ close to the operating sweep rate. The latter can be achieved by increasing the sheet resistance $\rho/d_n$ and/or decreasing the length of the twist pitch.

**B) Reduction of critical current by defects.**

The multifilament coated conductors inevitably will be much more susceptible to defects than the currently manufactured wide uniform tapes. The degradation of current-carrying capacity of the striated conductor by defects can be estimated in a model of uncorrelated defects. Let us consider the worst case scenario – just one defect is sufficient to block the current flow in a given stripe. The probability that a stripe of length $L$ has $n$ defects is given by the Poisson distribution

$$p_n = e^{-\bar{n}} \bar{n}^n / n!,$$

where $\bar{n} = L/\bar{l}$ and $\bar{l}$ is the average length of a stripe without a single defect. The probability that a striated tape comprised of N stripes has $k$ stripe unblocked and therefore conducting, and the rest are blocked by one or more defects is given by

$$P_k = \frac{N!}{k!(N-k)!} p_0^k (1-p_0)^{N-k}.$$

The average number of unblocked stripes $\bar{k} = N p_0 = N e^{-L/\bar{l}}$ and the average critical current

$$\bar{I}_c = I_c^{st} e^{-L/\bar{l}},$$

where $I_c^{st} = NJ_cW_n$ is the maximum critical current of the multifilament conductor. For example, if we adopt a requirement that $\bar{I}_c$ should not be less than a certain fraction $\varepsilon$ of $I_c^{st}$, the useful length of the conductor

$$L \leq \bar{l} \ln(1/\varepsilon) . \qquad (10)$$

This is a severe restriction because it dictates that the conductor as a whole is only as good as an individual stripe. For example, if one requires that $\bar{I}_c$ should not be less than 75% of $I_c^{st}$,

$$L \leq \bar{l} \ln(4/3) \approx 0.3\bar{l} .$$

The defect-free length of an individual stripe $\bar{l}$ can be estimated if we adopt a certain model of defects. In Refs. [26,27] such a model is based on the assumption that the grain misalignment is the main reason for reduction of critical current in the long tapes. The probability of appearance of a cluster of defects decreases exponentially with the size of the cluster. This leads to the logarithmic decrease of the critical current with length of the conductor [27]. However, other reasons such as equipment malfunction during the manufacturing - a drop of solvent or a speck of dust in the wrong place, etc. may lead to the same result. Without specifying the nature of defect, let us assume that the probability of a defect size greater than $R$ is

$$\kappa \approx e^{-\alpha R} .$$

The number of such defects $M$ in a given stripe increases linearly with the area of the stripe.

$$M \approx \frac{W_n l}{l_0^2} e^{-\alpha R} .$$

Here $l_0^2$ is the characteristic area determined by the nature of defects. The prefactor determines the number of places – "trouble spots" – in a given stripe where a *large defect* can nucleate. Models based on the assumption that grain misalignment is the main factor limiting the critical current in the long tapes consider every grain as a potential trouble spot and take $l_0$ to be of the order of the average grain size (see [26,27] and references therein). This may not necessarily be true. If the dominant source of defects is equipment malfunction (during manufacturing and/or exploitation) the characteristic area required for a large defect to appear may be unrelated to and, in practice, much greater than the grain size.

A stripe will be blocked if the size of the defect is comparable to the width of the stripe $R \approx W_n = W/N$. The defect-free length of an individual stripe is determined by the condition $M \approx 1$ which leads to

$$\bar{l} \cong \frac{l_0^2}{W_n} e^{\alpha W_n}. \tag{11}$$

Let us consider two extreme limits in which a long defect-free length can be achieved. First is a situation when the most prevalent defects simply have very few trouble spots on the tape where they can appear. But once a trouble spot is created, its probability to become a defect is of the order of unity. In this case the defect-free length is determined by the prefactor in (11):

$$\bar{l} \cong l_0^2 / W_n \; ; \; e^{\alpha W_n} \approx 1.$$

In the opposite limit the long defect-free length may result from low probability of a defect, even though the potential sites where they can appear are plentiful:

$$K \equiv W_n \bar{l} / l_0^2 \gg 1 \; ; \; e^{-\alpha W_n} \approx K^{-1} \ll 1.$$

Let $L_N$ be the useful length of a conductor comprised of N stripes. How will this length change when the number density of stripes doubles to 2N? According to Eqs. (10) and (11):

$$L_{2N} \propto \frac{2 l_0^2}{W_n} \left( e^{\alpha W_n} \right)^{1/2} = 2 L_N \left( \frac{l_0^2}{\bar{l}_N W_n} \right)^{1/2} = \frac{2 L_N}{K^{1/2}}. \tag{12}$$

In the first scenario, the useful length may not change much, because the subdivision of the tape into twice as many stripes does not increase the number of potential trouble spots, $K \approx 1$. In the second scenario, however, the effect of subdivision on useful length will be much more spectacular because the probability of a blockage of a narrower stripe increases exponentially:

$$\left( e^{-\alpha W_n} \right)^{1/2} \gg e^{-\alpha W_n}.$$

Thus, if the number of potential trouble spots along the defect-free length is large, $K \gg 1$, the doubling of the number of stripes decreases the useful length by a factor $K^{-1/2}$.

For example, let us consider the application of Eq. (12) to the models of grain misalignment[26,27]. In this case every grain has to be treated as a potential trouble spot. We know from experience that the useful length of a 20-filament conductor can exceed at least 10 cm[20,22]. Taking $\bar{l} = 10 \, cm$, $W_n = 0.5 \, mm$ and the average grain size $l_0 \approx 30 \, \mu m$ [28], we get $K \cong 5 \times 10^4$. The Eq. (12) would suggest then, that a 40-filament conductor will have negligible useful length. This, however, is not the case as shown below in section 3. The 40-filament, 10 cm long conductor is shown to have critical current of about 100 A, comparable to that of the 20-filament one. Thus, the number density of the trouble spots has to be much smaller than the number density of grains and, therefore, the nature of current limiting defects in coated conductors is not grain misalignment. This statement should be qualified by the fact that the width of the filament in the 20- and 40-filament experimental conductors is much greater than the grain size. In the future, multifilament conductors with the stripe width of the order of 50-100 μm the grain misalignment may become a more important factor limiting their reliability.

There is little doubt that in practical, meters long, multifilament conductors of the future, the problem of reduction of critical current by defects, summarized by Eq. (12), will be real and needs to be addressed. A way to alleviate this problem is not to segregate the stripes completely, but to leave a sparse network of narrow superconducting bridges that allow the current sharing between the stripes [23,27]. However, one has to ensure that the superconducting interconnections do not defeat the main purpose of striation – reduction of the magnetization losses. The main goal of this paper is to show that properly placed and spaced by the distance comparable to the twist pitch bridges do not increase substantially the power loss in such multiply connected superconducting films in comparison with the fully striated films.

**3. Power loss in fully striated (simply connected) conductors.**

All samples, fully striated and multiply connected ones, were cut from a 61 cm long single piece of coated superconductor provided by SuperPower Inc. [14]. The YBCO layer in this sample was deposited on buffered Hastelloy substrate 90 μm thick and covered with silver about 5μm thick. In this section we present the ac losses in two 1x10 cm samples that were divided into 20 and 40 stripes respectively without superconducting bridges between the stripes.

Striation of the tapes is accomplished by laser micromachining utilizing a frequency tripled diode-pumped solid-state Nd:YVO$_4$ laser at 355 nm wavelength. Details of striation by laser ablation are given in Ref. [29]. For striation of coated conductors we have used the maximum scanning speed of the laser beam of 150 mm/s. The length over which a sample could be scanned without moving it was approximately 2.5 cm. Thus, the 40-stripes, 2.5 cm long section can be processed in 8s. The translation speed of the moving stage which carried the sample was close to 17 mm/s. Therefore, it takes about 10s to process a 2.5 cm section and start a new one. The total rate of striation of the 40-filament/(cm width) conductor is 15 cm/minute. This figure should not be considered representative of the potential of laser ablation because we have used a system that was available and no effort was applied to maximize the throughput. The relative merits of using laser ablation for mass production of striated conductors can only be determined when the results of the competing approaches (photolithography, etc.) become available.

Figure 1(a) shows a small area of the 40-filament sample. The distance between the grooves is 250 μm. Figure 1(b) shows a typical profile of a groove. The electrical connection between the substrate and YBCO and Ag is provided by Hastelloy that was melted and splashed across the walls of the groove during ablation. The amount of YBCO lost as the result of ablation is about 6% and 12% for the 20-filament and 40-filament conductors respectively.

The magneto-optical (MO) imaging technique allows us to obtain an overall assessment of the quality of the superconducting film. MO characterization was performed using a Bi-doped magneto-optical garnet film with in-plane magnetization grown on Gadolinium Gallium Garnet substrates [30-32]. The sample was mounted on a cooling finger of a continuous flow optical cryostat capable of cooling to ~ 6K located on an X-Y stage of a polarized optical microscope in reflective mode. To register the normal component of the magnetic flux distribution on the sample surface, the indicator film was placed on the top sample face. The external magnetic field was applied perpendicular to the plane of the film by a small solenoid surrounding the cryostat-cooling finger. A silicon diode and a LakeShore temperature controller adjusted the sample temperature. A digital camera was used to record the magneto-optical images.

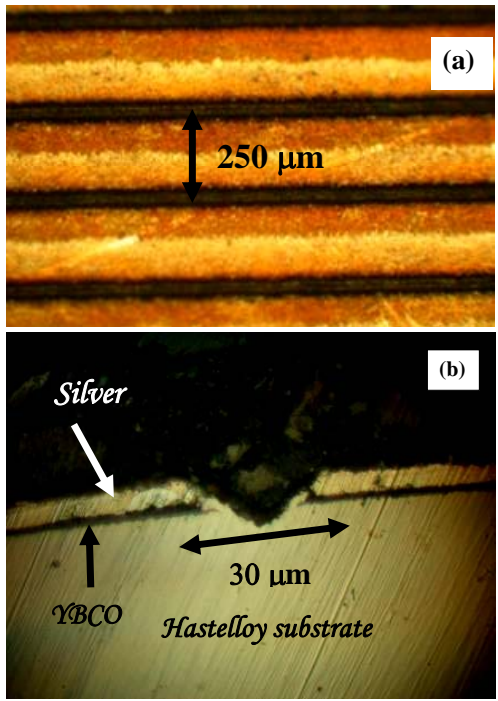

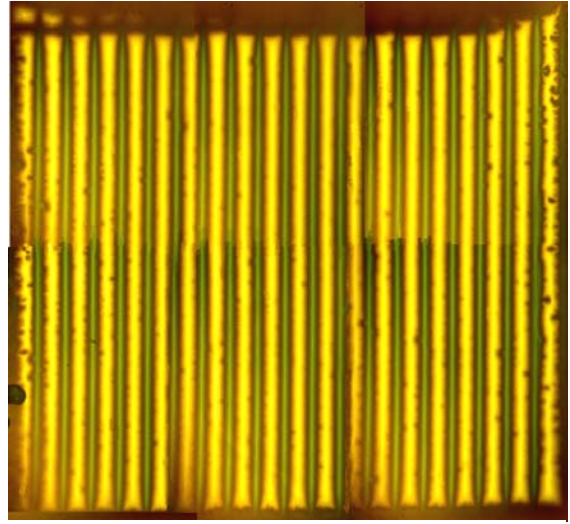

*Fig.1 (Color online) (a)* Shown is a small area of a 1 cm wide coated conductor divided into 40 stripes. The distance between the centers of the grooves is 250 μm. The non-superconducting grooves are about 30 μm wide. The width of a superconducting stripe is about 220 μm. The top layer is silver. *(b)* Profile of the groove; the depth is about 10 μm and it extends well into the substrate, cutting through silver, YBCO and buffer layers

*Fig. 2. (Color online)* Magneto-optical image of a fully striated 1x1 cm sample divided into 20 stripes. The sample was field-cooled in B=0.1 T to T=10 K and then the field was turned off. Bright (yellow online) bands correspond to magnetic flux trapped by the superconducting stripes. Dark (green) spaces are non-superconducting grooves.

Figure 2 shows a magneto-optical image of a 1x1 cm area of a sample divided into 20 stripes. The sample was field cooled in the applied field of 100 mT to 10 K and then the field was turned off. As the result, the magnetic flux was trapped in the superconducting stripes (bright bands, yellow online). This and other images below are taken in color (available in online edition) which allows to show the direction of the magnetic flux density. Yellow color indicates the trapped magnetic flux directed "up" – in the direction of the applied field. Green color indicates the stray field directed "down". Inside the grooves there are faint green lines indicating weak demagnetizing field caused by the supercurrents confined inside each stripe. The image demonstrates that laser ablation produces fairly uniform stripes of good quality, except at the edges where the superconducting film shows some fraying, which apparently was present even before striation.

In order to measure the magnetization losses, the samples were placed inside the bore of a dipole magnet that generated time-varying magnetic field. The entire system was cooled in liquid nitrogen. The magnetization loss was measured using a linked pick-up coil[20]. In the experiment, the frequency was varied from 11.3 to 171.0 Hz.

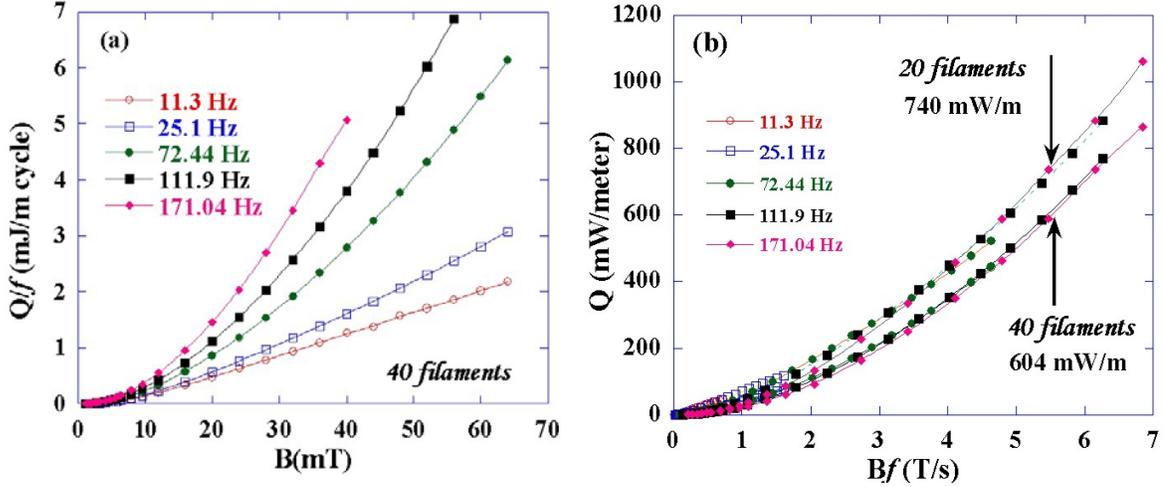

*Fig.3* (Color online) *(a)* Energy loss per meter length, per cycle for different frequencies in the 40-filament sample, plotted as a function of the amplitude of applied field. *(b)* the same data as in *(a)* presented as the power loss per meter length vs sweep rate. Also shown is the power loss in the 20-filament sample. For comparison, the values of the power loss in both samples at Bf = 5.5 T/s are indicated.

Figure 3 shows the ac losses in external time-varying magnetic field. The frequency range and the amplitude of the field are indicated in the figure. In Fig. 3(a) the losses in the 40-filament sample are shown in the traditional form as the energy loss per cycle. The per cycle loss increases with frequency and at all frequencies, except the lowest, the field dependence is non-linear indicating a substantial contribution of coupling losses.

The Fig. 3(b) shows the power loss vs sweep rate for both 20- and 40-filament samples. It demonstrates that the power loss depends on one variable – the sweep rate, see Eq. (8). Note that although the 40-filament sample has lower loss than the 20-filament one, the difference is not large. For example, at the sweep rate of 5.5 T/s the 20-filament sample dissipates 740 mW/m, while the 40-filament sample dissipates 604 mW/m, which is only about 20% less. The reason for that is that the coupling losses are dominant in this range of the sweep rate and further increase of the number density of stripes will not bring any substantial reduction of losses.

In order to separate the hysteresis and coupling losses it is convenient to present data as suggested by Eq. (8) and shown in Fig. 4 – as power loss per unit of the sweep rate [22]. As the field amplitude increases above the level of full penetration, the linear dependence of $Q/Bf$ on the sweep rate, given by Eq.(8), becomes obvious. The straight dashed lines in Fig.4 are obtained by fitting the 111.9 Hz data.

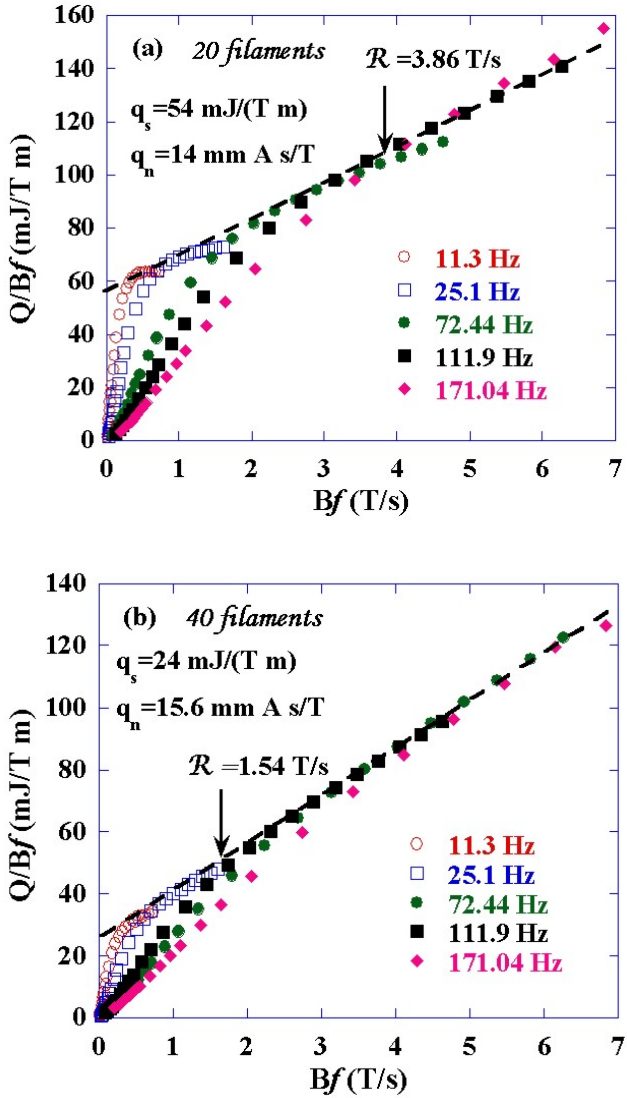

**Fig.4** (Color online) The same data as in Fig. 3(b) presented as power loss per unit of the sweep rate vs sweep rate (see Eq. (8)). (**a**) the 20-filament sample and (**b**) the 40-filament sample. For each sample the hysteresis loss $q_s$ (the y-axis intercept) and coupling loss $q_n$ (the slope) are shown and the break-even rate, $\mathcal{R}$=3.86 T/s and 1.54 T/s respectively, are indicated by arrows.

The intercept of the straight lines is the hysteresis loss per unit of the sweep rate denoted as $q_s$ in Eq. (8) and the slope is the coupling constant $q_n$. From experience[22] we know that the relationship $q_s = W_n I_c$ holds well. Thus, the bulk average critical current can be estimated from the values of $q_s$ shown in Figs. 4(a,b) and the known width of the filaments (note that 1 mJ/T m $\equiv$ 1 mm A). We take the width of the non-superconducting grooves to be close to 30 μm, Fig. 1(b). Therefore, assuming the width of the superconducting stripes to be 0.47 and 0.22 mm respectively and taking the values of $q_s$ = 54 and 24 mJ/Tm, we obtain $I_c$=115A for the 20-filament sample and $I_c$=109 A for the 40-filament one. This is consistent with the expected 6% and 12% reduction of the critical current in these samples due to removal of YBCO by ablation. The measured transport critical current in the 40-filament sample was close to 100 A[33].

The break-even rate defined by Eq.(9) can be estimated as follows:

$$\mathcal{R} \approx \frac{6}{\pi^2} \frac{q_s \rho}{L^2 d_n W} \quad (13)$$

The resistivity of Hastelloy is practically temperature independent, $\rho \approx$ 130 μΩ cm. The thickness of the substrate is close to 100 μm. Therefore, the sheet resistance of the substrate $\rho/d_n \approx$ 13 mΩ[20]. For L=10 cm and W=1 cm, we get $\mathcal{R} \approx$ 4.3 T/s for the 20-filament and $\mathcal{R} \approx$ 1.9 T/s for the 40-filament sample. The coupling constants $q_n = q_s / \mathcal{R}$ are equal to 14 mm A s/T and 15.6 mm A s/T respectively. The agreement with the experimental value is worse for the 40-filament sample, indicating that Eq.(13) gives a correct order of magnitude estimate, but misses a more subtle dependence of the coupling loss on the number of stripes.

## 4. Multiply connected conductors.

A way to alleviate the problem posed by small defects, as discussed above, is to allow current sharing between the stripes. Inherently, the interconnections lead to increased magnetization loss[34]. Here we address a question of how compatible is the presence of superconducting bridges between the current-carrying stripes with the requirement of low magnetization losses.

Figure 5 shows sketches of two tested patterns. In Fig. 5(a) the "brickwall" pattern is shown. The positions of bridges connecting a given stripe to the rest of the sample alternate and the coupling current, directed perpendicular to the stripes, has to take a meandering path. Figure 5(b) shows a different pattern – one continuous bridge connecting all stripes ("fishnet"). In either pattern the distance between the successive bridges is called the current transfer length $L_t$ [35].

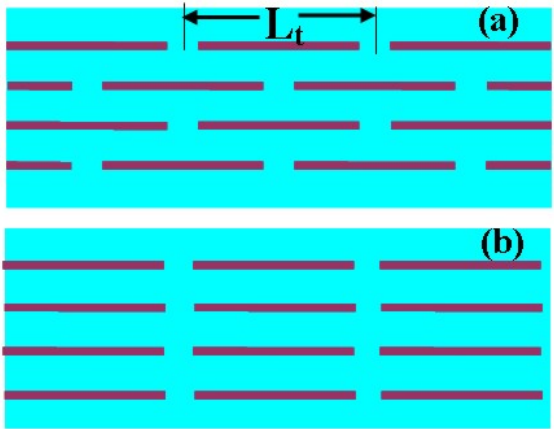

*Fig. 5 (Color online) Two multiply connected patterns of the superconducting layer that have been tested. The superconducting area is shown in gray (blue online); the non-superconducting cuts are shown by dark lines. (a) Alternating bridges (brickwall pattern). (b) parallel bridges (fishnet pattern). The distance between the successive bridges indicated by the arrow is the current-transfer length $L_t$.*

The minimum width of the bridge in both patterns should be half of the width of a stripe. This allows the transport current to flow in- and out of a given stripe through the two nearest to the damaged area bridges without raising the level of criticality $j/j_c$ in the bridge above the average in the conductor. Thus, a defect in a given stripe can block it only over a distance $L_t$. The bridges will allow the current to circumvent the damaged length leaving the rest of the stripe usable. The criterion for the length of $L_t$ is similar to the condition (10):

$$L_t \leq \bar{l} \ln(1/\varepsilon). \qquad (14)$$

**A) Fishnet pattern.**

Figure 6(a) shows a sketch of a sample with one bridge located in the center of the sample. Comparing with Fig. 5(b), we see that this is an imitation of a twisted conductor with the current transfer length equal to half of the twist pitch and the bridge located exactly halfway between the nodes of the twisted tape[27]. A sample of finite length L is equivalent to a twisted tape with half of the twist pitch equal to L. Thus, our 10 cm long samples imitate a twisted tape with 20 cm long twist pitch. In Fig. 6(b) the area around the bridge of the actual 20-filament sample is shown. The width of the bridge is about 200 μm, which is about half of the width of an individual stripe. The laser ablation was carried out similarly to the 20-filament fully striated sample whose losses are shown in Fig. 4(a).

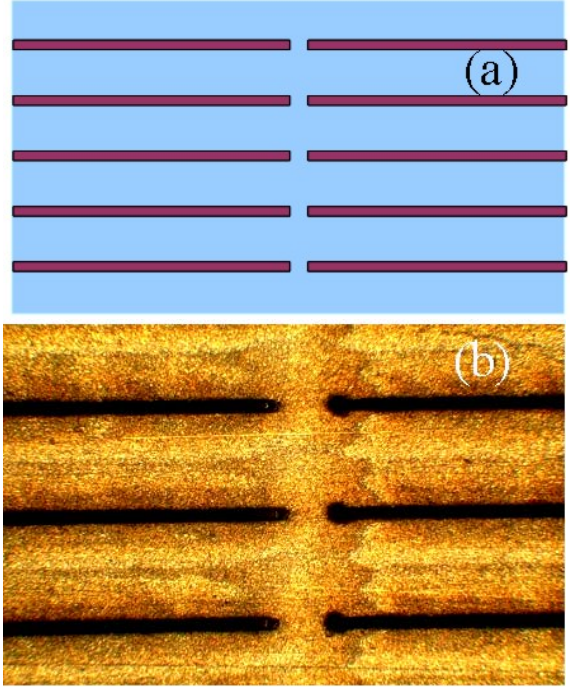

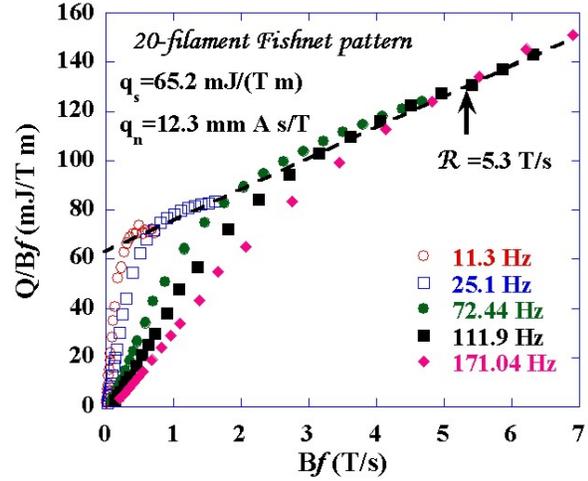

*Fig.6* (Color online) **(a)** Sketch of the multifilament sample with one superconducting bridge in the center, fragment of the fishnet pattern shown in Fig.5(b). **(b)** A small area of the actual 20-filament sample. The length of the sample is 10 cm, width 1 cm, width of the bridge (about 200 µm) is approximately half of the width of an individual stripe. The top layer is silver.

*Fig. 7* (Color online) Power loss per unit of the sweep rate vs sweep rate (see Eq. (8)) for the 20-filament sample with a bridge in the center as shown in Fig 6. The hysteresis loss $q_s$, the coupling loss $q_n$ and the break-even rate $\mathcal{R}$ = 5.3 T/s are indicated. The straight dashed line is the linear fit to 111.9 Hz data.

The location of the bridge in the center of the sample is the most favorable from the loss reduction point of view. The transverse electric field, Eq. (6), is minimum and the additional loss in the bridge can be roughly estimated [27] using Eq.(3) and treating the bridge as a stripe whose *length* is W and width is Δ. Thus, the amount of heat released in the bridge per unit length of the sample

$$Q_{br} \approx \frac{J_c \Delta^2 W B f}{L} \equiv I_c W_n B f \frac{\Delta^2}{W_n L}. \quad (15)$$

If we take $\Delta = W_n/2$, the contribution of the centrally located bridge to losses is negligible in comparison with that of the stripes, Eqs. (4) and (8). Moreover, the width of the bridge can be increased to be equal or even greater than $W_n$ without paying significant penalty in terms of increased losses. Increasing Δ up to $\Delta \approx (W_n L)^{1/2}$ will only double the hysteresis loss. Wider bridge will be a more reliable means for mixing and redistributing the transport current between the stripes.

Figure 7 shows the losses in the sample with one central bridge, shown in Fig 6. The hysteresis constant $q_s$ is somewhat higher, and the coupling constant $q_n$ is somewhat lower than in the simply connected sample, Fig. 4(a), but the total losses in the fully striated and fishnet pattern are almost the same as is illustrated below. At this point the difference between the characteristic constants describing the losses may be attributed to the sample-to-sample variations of the critical current.

**B) Brickwall pattern.**

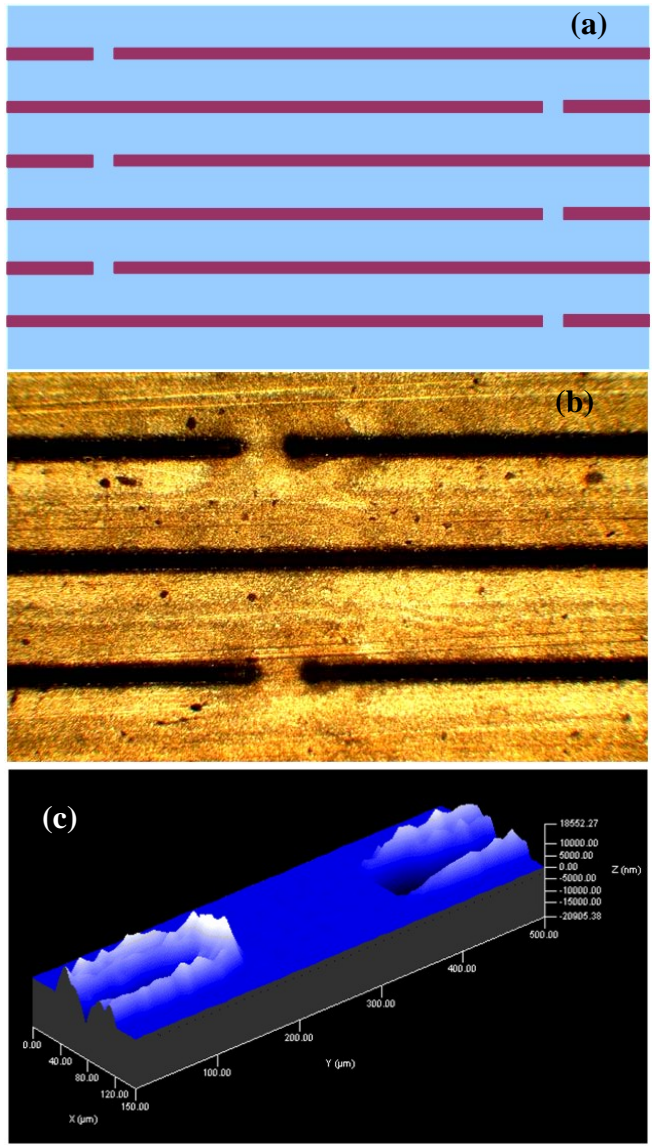

Figure 8(a) shows a sample whose pattern of bridges is part of the brickwall pattern shown in Fig. 5(a). In the 10 cm long sample the bridges are placed 1 cm from the nearest edge. This corresponds to the current transfer length 16 cm in the twisted conductor with the twist pitch 20 cm[27]. Figure 8(b) is a photograph of the actual 20-filament sample. Figure 8(c) shows a profile of the YBCO layer in the vicinity of a bridge measured by profilometer after the cap silver layer was etched away.

Figure 9(a) shows a magnified image of the YBCO layer near one of the bridges. The distance between the grooves is 500 μm, the width of the bridge connecting the neighboring stripes is about 200 μm. Figure 9(b) shows a photograph of exposed YBCO layer that was patterned as shown in Fig. 5(a) with the current transfer length $L_t$ =3300 μm, much smaller than that in Fig. 8(a).

Fig. 10 shows magneto-optical images (Figs. 10(a) and (c)), and a conventional optical photograph (Fig. 10(b)) of the surface of a 1x1 cm sample with the same pattern of cuts as that shown in Figs. 5(a) and 9(b). The MO pictures shown in this figure were taken at 77 K, the same temperature at which the losses were measured. The sample was cooled in zero magnetic field and then the field was slowly increased in steps of 4mT to a maximum of 40 mT and then reduced back to zero. This procedure roughly imitates half a cycle of an alternating applied magnetic field. Figure 10(a) shows the distribution of the flux density at the start of this "cycle" when a small external field of 4 mT was applied. The bright areas identify the maximum flux density penetrating through the grooves. The photograph shows three rows of bridges. The middle row is indicated by the double arrows pointing at the same area in all three photographs. The bridges are superconducting and partially prevent the magnetic flux from entering the central area of the grooves.

*Fig. 8 (Color online) (a) A sketch of the sample with alternating bridges (brickwall pattern). The bridges are placed 1 cm away from the nearest edge. (b) A small area of the actual 20-filament sample. The top visible layer is silver. The width of the bridge (about 200 μm) is approximately half of the width of an individual stripe. (c) Profile of the YBCO layer around the bridge. The cap silver layer was etched away.*

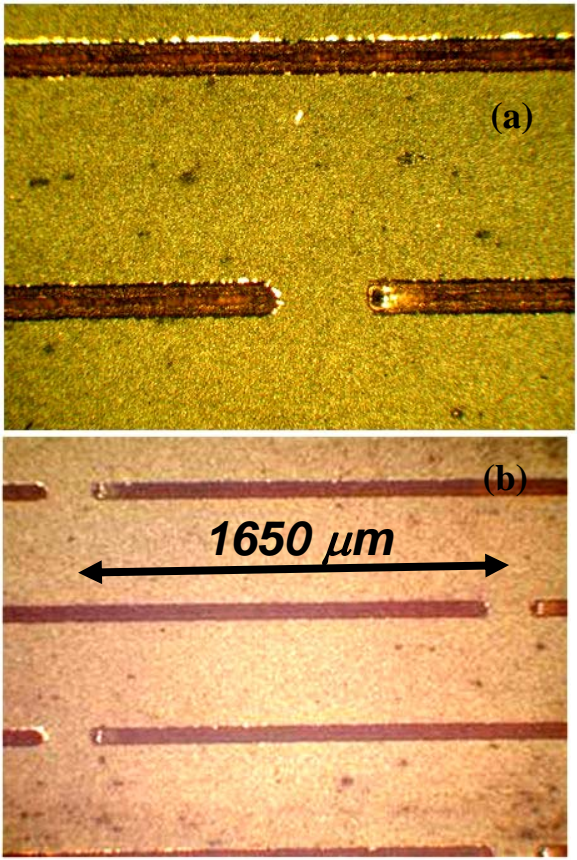

*Fig. 9 (Color online) (a) In this photograph a bridge between two stripes in exposed YBCO layer is shown. The width of the stripes is about 500 μm, the bridge is about 200 μm wide. (b) The exposed YBCO layer cut in the brickwall pattern with the transfer length 3300 μm.*

The arrows at the bottom of the picture point to the grooves. One can see a big difference in the flux density in odd- (1,3, …) and even-numbered (2,4, …) grooves. This is because the even-numbered grooves extend to the edge of the sample and external field penetrates freely through them. Each of the odd-numbered grooves is crossed by two additional bridges, located above and below the photograph frame. These bridges impede the penetration of magnetic flux into and its withdrawal from these grooves to a greater degree than that in the even-numbered grooves.

Figure 10(c) shows the flux distribution at the end of the "half-cycle" when the external field is turned off after reaching the maximum value of 40 mT. The bright lines show the multiply connected "backbone" of the trapped magnetic flux and the characteristic rooftop pattern in each of the superconducting stripes[12]. The bridges trap magnetic flux as well. The stray magnetic field in the grooves is smaller than in the stripes and, therefore, the grooves appear as the darker contrasted areas. When the applied field decreases, the odd-numbered grooves retain a greater flux density and appear brighter than the even-numbered ones that extend to the edge of the sample. One can also see the uneven quality of the bridges. One of them (encircled areas) is non-superconducting and therefore invisible in the MO images. Others display different degrees of field penetration.

This and other images of the same sample taken under different field conditions show that the superconducting bridges create a gradient of the magnetic flux density along the groove and, therefore, induce the current flowing across the bridges which contributes to dissipation. However, it also shows that the bridges are capable of redistributing the supercurrent between the stripes when needed.

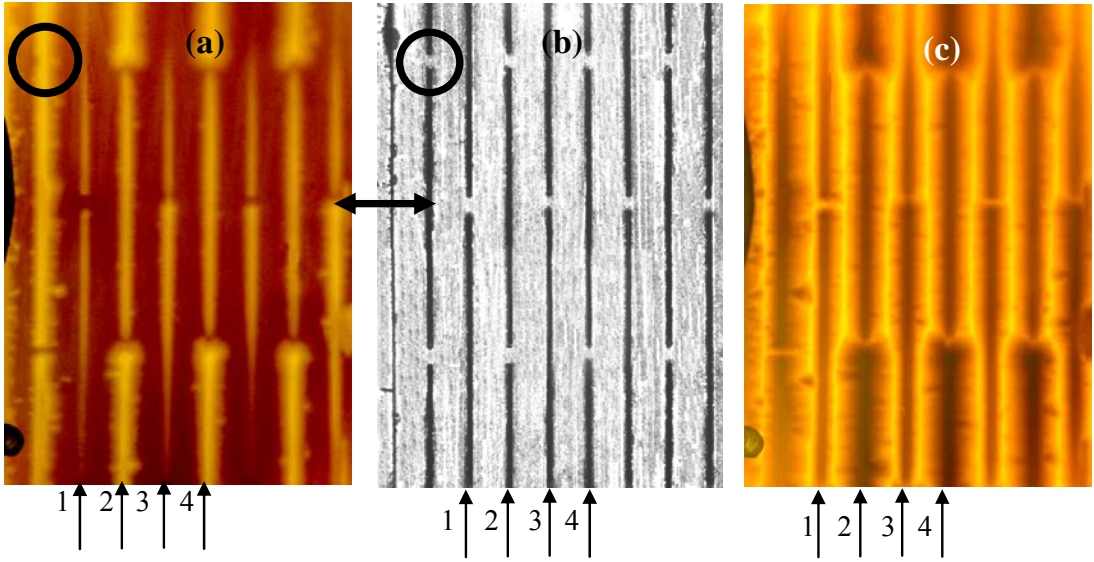

*Fig. 10* (Color online) Images of the 20 stripes/cm film with alternating bridges (same pattern as in Fig. 9(b)). The sample was cooled in zero field to 77 K and then the magnetic field was applied in steps (increased to a maximum of 40 mT and then decreased back to zero). (**a**) MO image in external field 4 mT at the start of the cycle. Bright areas correspond to maximum flux density. The arrows (1, 2, 3, 4 …) point at non-superconducting grooves. (**b**) Optical photograph of the same area. Double arrows indicate one of the three visible rows of bridges in each figure. The distance (along a groove) between successive bridges $L_t$ = 3300 μm. (**c**) MO image of the same area as in (**a**) after the external field is turned off (end of half-cycle). Non-superconducting (defective) bridge is encircled.

The electric field in the area of the bridges is given by Eq.(6). Using Eq.(1) the additional loss associated with a single bridge located at a distance $x$ from the centerline can be estimated as follows:

$$Q_{br}^{(1)} \approx \frac{2}{\pi} J_c B \omega x \int dA_s \qquad (16)$$

Here integration is carried out over the effective area of the bridge. Looking at Fig. 10(a,c) one can guess that the effective length of this area is comparable to the width of the groove and certainly less than the width of a stripe: $A_s \approx \lambda_b \Delta$ with $\Delta \leq \lambda_b \leq W_n$. In the sample shown in Fig. 8(a) there is one bridge per groove, N-1 bridges total. The total additional loss associated with bridges per unit length of the sample is given by

$$Q_{br} \approx I_c B f W_n \left( \frac{4x}{L} \frac{\lambda_b \Delta}{W_n^2} \right). \qquad (17)$$

In the sample shown in Fig. 8(a) $x=4cm$, $L=10cm$, $\Delta/W_n \approx 0.5$ and if we take the upper limit for $\lambda_b/W_n \approx 1$, the hysteresis loss should increase by 80% in comparison with the contribution of stripes, Eqs. (4).

Figure 11 shows the power loss in the sample patterned as shown in Fig. 8(a). The hysteresis loss $q_s$ is about 20% higher than in fully striated sample, Fig. 4(a). This is consistent with the estimate (17) where $\lambda_b/W_n \approx 1/4$. The coupling constant $q_s$ is practically the same as in fully striated sample, Fig. 4(a).

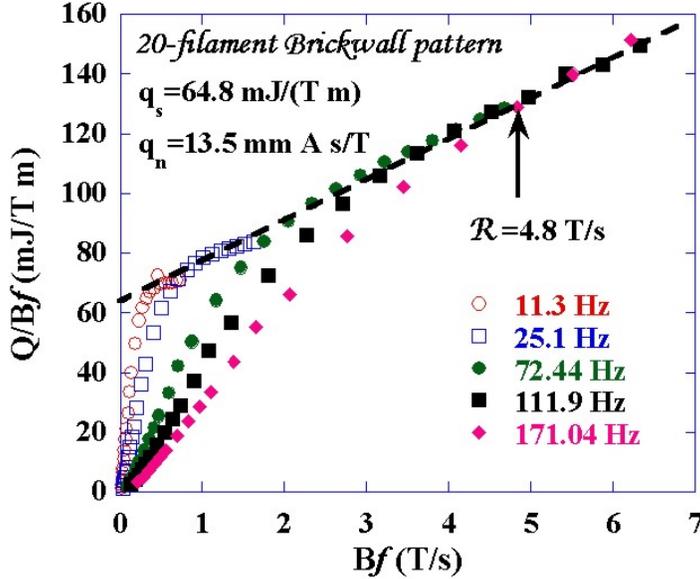

*Fig. 11 (Color online) Power loss per unit of the sweep rate vs sweep rate (see Eq. (8)) for the 20-filament sample with alternating bridges as shown in Fig 8(a). The hysteresis loss $q_s$, the coupling loss $q_n$ and the break-even rate $\mathcal{R} = 4.8$ T/s are indicated. The straight dashed line is the linear fit to 111.9 Hz data.*

Figure 12 shows the 111.9 Hz data for all three tested patterns. The brickwall pattern clearly has higher losses due to higher hysteresis loss. The loss in the sample with the one centrally located bridge is closer to that in the fully striated sample.

## 5. Summary.

The magnetization loss measurements demonstrate that multiply connected superconducting tapes in which the individual current-carrying stripes are connected with each other by a sparse network of superconducting bridges can have the total loss comparable to that of the fully striated tapes. The allowable, by condition of loss reduction, linear density of bridges (along the filaments) depends on the length of the twist pitch, and should not exceed 1-4 bridges per half of the twist pitch. Without current sharing the multifilament conductors will be very difficult to manufacture in substantial length because of the cumulative effect of relatively small defects and defects clusters on its current-carrying capacity. This will particularly be true in conductor with very fine filaments. We should note that the effectiveness of the superconducting bridges in increasing the useful length of the multifilament conductor has not been experimentally confirmed yet. The results presented here indicate that if such interconnections will prove to be effective and necessary in increasing the reliability of the future coated conductors, the increase in magnetization losses by interconnections is tolerable.

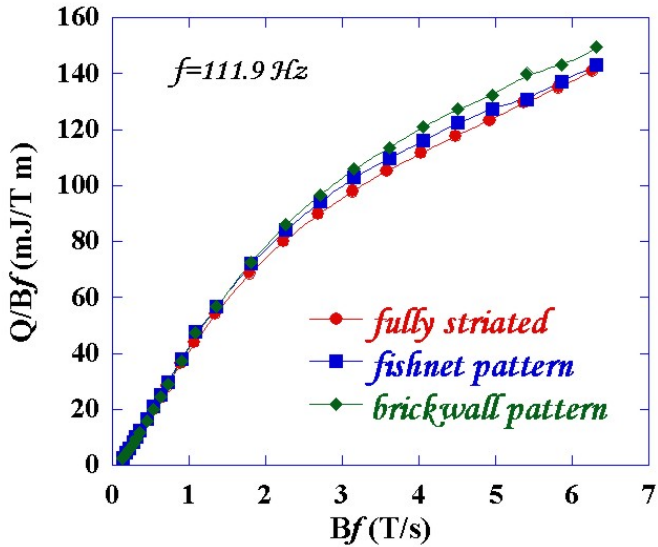

*Fig. 12 (Color online) For comparison, the data for three 20-filament samples with different patterns - fully striated (Fig. 4(a)), fishnet (Fig. 7) and brickwall (Fig. 11) are shown together. To avoid clutter, only one frequency, 111.9 Hz, is included.*

Another noteworthy point is that further reduction of magnetization losses, beyond what has been achieved in the 20-filament conductors, cannot be accomplished without solving the problem of coupling losses. As the data in Fig. 3(b) demonstrate, further increase in the number density of stripes brings rapidly diminishing results because the main contribution to losses comes from the coupling currents. Since the twist pitch length cannot be decreased substantially below 5-10 cm, other solutions have to be found for reducing the coupling losses. These will have to lead to drastic increase in the effective resistance of the grooves by at least one or two orders of magnitude over that achieved in our samples.

**Acknowledgements:**
We would like to thank M. Sumption and D. Larbalestier for very useful conversations and discussions, and J. Murphy, J. Kell, and N. Yust for technical assistance. At YNU this work was partially supported by the US Air Force Office of Scientific Research under contract AOARD-03-4031. The work at UW Madison has been supported by an AFOSR Multidisciplinary University Research Initiative.  One of the authors, G.A.L, was supported by the National Research Council Senior Research Associateship Award at the Air Force Research Laboratory.

# References

[1] S. S. Kalsi, K. Weeber, H. Takesue, C. Lewis, H.-W. Neumueller, and R. D. Blaugher, Proc. IEEE 92, 1688 (2004)
[2] D. U. Gubser, Physica C **392-396**, 1192 (2003).
[3]  P. N. Barnes, G. L. Rhoads, J. C. Tolliver, M. D. Sumption, and K. W. Schmaeman, IEEE Trans. Mag. 41, 268 (2005); P.N. Barnes, M.D. Sumption, and G.L. Rhoads, Cryogenics (in press).
[4] B. Oswald, K.-J. Best, T. Mayer, M. Soell and H. C. Freyhardt, Supercond. Sci. Technol. 17, S445 (2004).
[5] W. J. Carr, AC Loss and Macroscopic Theory of Superconductors, 2$^{nd}$ edition, Taylor and Francis, New York (2001).
[6] S. P. Ashworth, M. Maley, M. Suenaga, S. R. Foltyn, and J. O. Willis,  J. Appl. Phys. **88,** 2718 (2000)
[7] R. C. Duckworth, J. R. Thompson, M. J. Gouge, J. W. Lue, A. O. Ijadoula, D. Yu, and D. T. Verebelyi, Transport AC Loss Studies of YBCO Coated Conductors with Nickel Alloy Substrates, Supercon . Sci. Technol. **16**, 1294 (2003).
[8] A. O. Ijaduola, J. R. Thompson, A. Goyal, C. L. H. Thieme, and K. Marken, Physica C  **403,** 163 (2004).
[9] J. Ogawa, H. Nakayama, S. Odaka and O. Tsukamoto, Physica C 412-414,  1021 (2004).
[10] M. Iwakuma  *et al*. Physica C 412-414, 983 (2004).

[11]  T. Nishioka, N. Amemiya, Z. Jiang, Y. Iijima, T. Saitoh, M. Yamada and Y. Shiohara,  Physica C 412-414, 992 (2004).
[12] D. Larbalestier, A. Gurevich, D. M. Feldman, and A. Polyanskii, Nature 414, 368 (2001)
[13] M. W. Rupich *et al*.  IEEE Trans. Appl. Supercond. **13** 2458 (2003).
[14] V. Selvamanickam, H.G. Lee, Y. Li, X. Xiong, Y. Qiao, J. Reeves, Y. Xie, A. Knoll, and K. Lenseth, Physica C **392-396**,  859 (2003).
[15] T. Watanabe, Y. Shiohara, and T. Izumi, IEEE Trans. Appl. Supercond. **13,** 2445 (2003).
[16] A. Usoskin et al.  IEEE Trans. Appl. Supercond. **13,**  2452 (2003).
[17] W.J. Carr, and C.E. Oberly, IEEE Trans. on Appl. Supercond. **9**, 1475  (1999).
[18] B. A. Glowacki, M. Majoros, N. A. Rutter, and A. M. Campbell,  Cryogenics 41, 103 (2001)
[19] C.B. Cobb, P.N. Barnes, T.J. Haugan, J. Tolliver, E. Lee, M. Sumption, E.Collings, and C.E. Oberly, Physica C, **382**, 52 (2002).
[20] N. Amemiya,  S. Kasai, K.Yoda, Z. Jiang,  G. A. Levin ,  P. N. Barnes , and C. E. Oberly, Supercond. Sci. Technol. 17, 1464 (2004).
[21] M.D. SumptionT, E.W. Collings, and P.N. Barnes, Supercond Sci. Technol., **18**, 122-134 (2005).
[22] G. A. Levin, P. N. Barnes, N. Amemiya, S. Kasai, K.Yoda, and Z. Jiang,  Appl. Phys. Lett. 86, 072509 (2005).
[23] P. N. Barnes and M. D. Sumption, J. Appl. Phys. 96, 6550 (2004).
[24] E. H. Brandt and M. Indenbom, Phys. Rev. B 48, 12893 (1993)
[25] Y. Mawatari, Phys. Rev. B 54, 13215 (1996)
[26] N. Rutter and A. Goyal, in Studies of High Temperature Superconductors, pp. 377-398,  Springer, New York, 2004.
[27] G. A. Levin and P. N. Barnes, IEEE Trans. Appl. Supercond. 15, 2158 (2005).
[28] A. Goyal *et al*. J. Mater. Res. 12, 2924 (1997).
[29] K. E. Hix, M. C. Rendina, J. L. Blackshire, and G. A. Levin, cond-mat/0406311
[30] L.A. Dorosinskii, M.V. Indenbom, V.I. Nikitenko, Yu.A. Ossip'yan, A.A. Polyanskii, V.K. Vlasko-Vlasov,  Physica C 203 (1992) 149-156.
[31] A.A. Polyanskii, X.Y. Cai, D.M. Feldmann, D.C. Larbalestier, Nano-crystaline and Thin Film Magnetic Oxides (NATO Science Series 3. High Technology-Vol. 72), Edited by I. Nedkov and  M. Ausloos, p. 353-370. 1999 Kluwer Academic Publishers.
[32] A.A. Polyanskii, D.M. Feldmann and D.C. Larbalestier, Magneto-Optical Characterization Techniques. Chapter C3.4, Handbook of Superconducting Materials. Ed. D. Cardwell, University of Cambridge, UK; D. Ginley, NREL. IOP publishing Ltd 2003, p. 1551-1567.
[33] N. Amemiya,  K. Yoda, S. Kasai,  Z. Jiang,  G. A. Levin ,  P. N. Barnes, and C. E. Oberly. IEEE Trans. Appl. Supercond. 15, 1637 (2005).
[34] M. Majoros, B. A. Glowacki, A. M. Campbell,  G. A. Levin, P. N. Barnes, and M. Polak, IEEE Trans. Appl. Supercond. 15, 2819 (2005).
[35]  M.D. Sumption, Physica C, **261**, 245-258 (1996).